\documentclass[aps,prb,preprint,showpacs,superscriptaddress]{revtex4-1}
\pdfoutput=1 
\usepackage{times}
\usepackage{ifpdf}
\usepackage{graphicx}
\usepackage[T1]{fontenc}
\begin{document}
\title{Pseudorandom selective excitation in NMR}
\author{Jamie D. Walls}
\email{jwalls@miami.edu} \affiliation{Department of Chemistry,
 University of Miami, Coral Gables, FL 33124}
 \author{Alexandra Coomes}
\date{\today}
\begin{abstract}
In this work, average Hamiltonian theory is used to study selective excitation in a spin-1/2 system evolving under a series of small flip-angle $\theta-$pulses $(\theta\ll 1)$  that are applied either periodically [which corresponds to the DANTE pulse sequence] or aperiodically. First, an average Hamiltonian description of the DANTE pulse sequence is developed;  such a description is determined to be valid either at or very far from the DANTE resonance frequencies, which are simply integer multiples of the inverse of the interpulse delay.   For aperiodic excitation schemes where the interpulse delays are chosen pseudorandomly, a single resonance can be selectively excited if the $\theta$-pulses' phases are modulated in concert with the time delays.  Such a selective pulse is termed a pseudorandom-DANTE or p-DANTE sequence, and the conditions in which an average Hamiltonian description of p-DANTE is found to be similar to that found for the DANTE sequence.  It is also shown that averaging over different p-DANTE sequences that are selective for the same resonance can help reduce excitations at frequencies away from the resonance frequency, thereby improving the apparent selectivity of the p-DANTE sequences.  Finally, experimental demonstrations of p-DANTE sequences and comparisons with theory are presented.
\end{abstract}
%\date{07/17/2000}
\maketitle
\section{Introduction}
 Of the multitude of radiofrequency (RF) schemes used for exciting and controlling spin dynamics in NMR, most can be placed into one of two categories:  aperiodic RF pulse sequences or periodic RF pulse sequences.  For many aperiodic sequences, the RF phases, amplitudes and pulse delays are often chosen randomly or in a pseudorandom manner.  Such sequences have been used to generate white noise or broadband excitation in NMR noise spectroscopy\cite{Ernst70,Kaiser70,Kaiser74,Bartholdi76}, while sequences that generate colored noise have been used in early spin decoupling schemes, such as in noise decoupling\cite{Ernst66}.  Theoretical models of a spin system's response to pseudorandom pulse sequences typically use a Volterra or perturbation series in the randomly applied RF pulses\cite{blumich87}. Since many pseudorandom sequences are designed by considering only the first term in the Volterra series, pseudorandom sequences are typically low power and result in small, linear spin excitations.

Unlike aperiodic sequences, periodic RF pulse sequences are commonly used in a variety of NMR experiments and are often found to be superior to their pseudorandom counterparts; for example,  two periodic sequences, MLEV\cite{Levitt81} and WALTZ-16\cite{Shaka83}, provide better heteronuclear decoupling over noise decoupling under most conditions.  Many periodic RF pulse sequences are designed using average Hamiltonian theory (AHT)\cite{Haeberlen68}, where the necessary RF pulse sequence that generates a desired average Hamiltonian $\overline{H}_{avg}$ over a time $\tau_{c}$ must be determined ($\tau_{c}$ is the length of the pulse sequence).  Repeated application of the pulse sequence introduces frequencies into the dynamics that are integer multiples of $\frac{1}{\tau_{c}}$, which may result in higher-order contributions to $\overline{H}_{avg}$ that degrade the sequence's performance.  It has been previously noted that random or asynchronous pulse imperfections placed into pulse sequences can often improve their performance\cite{Bosman04}.  Recently, Uhrig dynamical decoupling (UDD) sequences\cite{Uhrig09} which utilize unequally spaced $\pi-$pulses, were shown to be superior in preserving spin coherence to the standard Carr-Purcell-Meiboom-Gill (CPMG) sequence\cite{Carr54}, which uses equally spaced $\pi-$pulses.

Selective pulses\cite{Freeman91} are one class of pulses that do not fall neatly into either category.  The design of most commonly used selective pulses, such as the gaussian and the sinc pulse shapes, is guided by the fact that a spin system's response to an applied pulse as a function of frequency/offset is proportional to the Fourier transformation of the applied pulse\cite{Tomlinson73}.  Using linear response to design selective pulses has been used to develop colored noise sequences for selective excitation in imaging applications\cite{Ordidge87}.  While sequences designed from the linear response are valid for small flip-angles, these pulses fail as the degree of excitation increases.  As such, most methods for designing selective pulses of arbitrary flip-angles use the linear response pulse shapes as starting points in numerical searches\cite{Veshtort04}.  However, pulse shapes generated by these numerical techniques often do not lend themselves to an easy physical interpretation behind their selectivity.

  One of the earliest and most easily understood periodic selective pulses that is rigorously valid for all flip-angles is the DANTE sequence\cite{Bodenhausen76}, which is shown in Fig. \ref{fig:figure1}(A).  The DANTE sequence consists of a series of $N$ small-tip, broadband $\theta$-pulses that selectively rotate those spins resonating at integer multiples of the interpulse delay by $\Theta=N\theta$ about an axis in the transverse plane.  The DANTE sequence's periodicity is responsible for this frequency response, which can be calculated analytically\cite{Canet95}.  To excite a single frequency, however, the periodicity of the DANTE sequence must be violated.  Breaking DANTE's symmetry for removing excitation at other frequencies has been previously accomplished by modulating the phases\cite{Kaczynski92}, amplitudes, and delays of the $\theta-$pulses\cite{Roumestand00}.  However, these excitation sculpting modifications of the DANTE sequence are still based on the assumptions of linear response.

In the following paper, we use AHT to provide insight into the selective excitation of a spin-1/2 system by a series of periodically and aperiodically small-flip $\theta$-pulses.  First, the conditions where an AHT description of the DANTE pulse sequence is valid is determined.  Next, an AHT description for a series of aperiodically spaced and phase-modulated $\theta-$pulses is developed.  Such sequences are referred to as pseudorandom-DANTE or p-DANTE selective pulses [Fig. \ref{fig:figure1}(B)].   Finally, experiments performed in acetone and in an acetone/DMSO/water solution are used to demonstrate and validate the selectivity of the p-DANTE sequences.
\section{General Theory}
\begin{figure}%[b!]
\includegraphics*[height=10cm,width = 8cm]{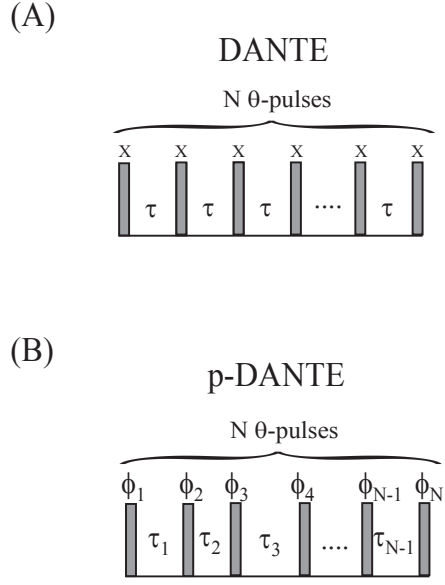}
{\small{\caption{Pulse sequence for (A) DANTE\cite{Bodenhausen76} and (B) pseudorandom-DANTE or p-DANTE selective excitation.  (A)  The DANTE sequence
 consists of a series of $N$ small-flip, $\theta$-pulses, equally spaced between periods of free evolution of time $\tau$.   The periodicity of the sequence
 results in a rotation of $\Theta=N\theta$ about an axis perpendicular in the transverse plane for those spins resonating (in the rotating frame) at a frequency $\nu_{Z}=\frac{n}{\tau}$ where $n$ is an integer.
   (B) The p-DANTE sequence consists of a series of $N$ unequally spaced small-flip, $\theta$-pulses where $\tau_{k}$ is the time delay between the $k^{th}$ and $(k+1)^{th}$, and $\phi_{k}=2\pi\sum_{k=1}^{k-1}\nu_{0}\tau_{k}$ is the phase of the $k^{th}$ pulse with $\phi_{1}=0$.  In the p-DANTE sequence, only those spins resonating (in the rotating frame) at a frequency given by $\nu_{0}$ are
   rotated by an angle $\Theta=N\theta$.  \label{fig:figure1}}}}
\end{figure}

Both the DANTE [Fig. \ref{fig:figure1}(A)] and the p-DANTE [Fig. \ref{fig:figure1}(B)] pulse sequences involve the application of a series of $N$ small flip-angle $\theta-$pulses that selectively rotate spins about an axis lying in the transverse plane by an angle $\Theta=N\theta$.  For the DANTE sequence, spins resonating at $\nu_{Z}=\frac{n}{\tau}$ are selectively rotated by $\Theta$ (where $n$ is an integer), whereas for the p-DANTE sequence, only those spins resonating at $\nu_Z=\nu_{0}$ are rotated by $\Theta$.
To understand the selectivity of both the DANTE and p-DANTE sequences within the framework of AHT, it is useful to revisit the dynamics of a
spin-1/2 system under a non-resonant RF irradiation.  The Hamiltonian during the
application of an RF pulse is given by: $\frac{\widehat{H}}{\hbar}=\omega_{z}\widehat{I}_{Z}+\omega_{RF}\left(\widehat{I}_{X}\cos(\phi)+\widehat{I}_{Y}\sin(\phi)\right)$,
where $\phi$ and $\omega_{RF}$ are
the phase and amplitude of the RF pulse, and $\omega_{Z}=2\pi\nu_{Z}$ is the resonance offset that the spin experiences in the rotating frame.
 The propagator for an RF pulse applied for a time $T_{p}$ can be written as:
\begin{eqnarray}
\widehat{P}^{\text{exact}}_{\phi}(T_{p})&=&\exp\left(-i\frac{\widehat{H}}{\hbar}T_{p}\right)=\exp\left(-i\left[\omega_{Z}\widehat{I}_{Z}+\omega_{RF}\left(\widehat{I}_{X}\cos(\phi)+\widehat{I}_{Y}\sin(\phi)\right)\right]T_{p}\right)\nonumber\\
&=&\cos\left(T_{p}\frac{\sqrt{\omega_{Z}^{2}+\omega_{RF}^2}}{2}\right)\widehat{1}-i\frac{2\sin\left(T_{p}\frac{\sqrt{\omega_{Z}^{2}+\omega_{RF}^2}}{2}\right)}{\sqrt{\omega_{Z}^{2}+\omega_{RF}^{2}}}\left(\omega_{Z}\widehat{I}_{Z}+\omega_{RF}\left(\cos(\phi)\widehat{I}_{X}+\sin(\phi)\widehat{I}_{Y}\right)\right)\nonumber\\
\label{eq:pulexact}
\end{eqnarray}

Alternatively, the propagator in Eq. (\ref{eq:pulexact}) can be transformed into
an interaction frame defined by $\omega_{Z}\widehat{I}_{Z}$ and is given by:
\begin{eqnarray}
\widehat{P}^{\text{exact}}_{\phi}(T_{p})&=&\exp\left(-i\omega_{Z}T_{p}\widehat{I}_{Z}\right)\widehat{T}\exp\left(-i\int^{T_{p}}_{0}\text{d}t'\frac{\omega_{RF}}{2}\left[\widehat{I}_{+}e^{i\left(\omega_{Z}t'-\phi\right)}+\widehat{I}_{-}e^{-i\left(\omega_{Z}t'-\phi\right)}\right]\right]\nonumber\\
&=&\widehat{U}_{\text{free}}(\omega_{Z}T_{p})\widehat{T}\exp\left(-i\int^{T_{p}}_{0}\text{d}t'\frac{\widehat{H}_{INT}(t')}{\hbar}\right)
\label{eq:pulprop}
\end{eqnarray}
where $\widehat{T}$ is the Dyson time-ordering operator, $\widehat{U}_{\text{free}}(\omega_{Z}T_{p})=\exp\left(-i\omega_{Z}T_{p}\widehat{I}_{Z}\right)$, and $\widehat{H}_{INT}(t')$, the Hamiltonian in the interaction frame, represents a purely phase-modulated RF pulse.   The time-dependent propagator in Eq. (\ref{eq:pulprop}) can be approximated by:
\begin{eqnarray}
\widehat{T}\exp\left(-\frac{i}{\hbar}\int^{T_{p}}_{0}\text{d}t'\widehat{H}_{INT}(t')\right)=\exp\left(-\frac{iT_{p}}{\hbar}\overline{H}_{p,\phi}\right)
\label{eq:pulprop2}
\end{eqnarray}
In Eq. (\ref{eq:pulprop2}), $\overline{H}_{p,\phi}$ is the average Hamiltonian\cite{Haeberlen68} and is given by $\overline{H}_{p,\phi}=\sum_{n=1}^{\infty}\overline{H}_{p,\phi}^{(n)}$ where the first two terms in the series are:
\begin{eqnarray}
\frac{\overline{H}_{p,\phi}^{(1)}}{\hbar}&=&\frac{1}{T_{p}}\int^{T_{p}}_{0}\text{d}t'\widehat{H}_{INT,\phi}(t')\nonumber\\
&=&\frac{\omega_{RF}}{2}\text{sinc}\left(\frac{\omega_{Z}T_{p}}{2}\right)\left(\widehat{I}_{+}e^{i\left(\frac{\omega_{Z}T_{p}}{2}-\phi\right)}+I_{-}e^{-i\left(\frac{\omega_{Z}T_{p}}{2}-\phi\right)}\right)\nonumber\\
\frac{\overline{H}_{p,\phi}^{(2)}}{\hbar}&=&\frac{1}{2iT_{p}}\int^{T_{p}}_{0}\text{d}t'\int^{t'}_{0}\text{d}t''\left[\widehat{H}_{INT,\phi}(t'),\widehat{H}_{INT,\phi}(t'')\right]\nonumber\\
&=&\frac{\omega_{RF}^{2}}{4it_{p}}\widehat{I}_{Z}\int^{T_{p}}_{0}\text{d}t'\int^{t'}_{0}\text{d}t''\left(e^{i\omega_{Z}(t'-t'')}-e^{-i\omega_{Z}(t'-t'')}\right)\nonumber\\
&=&\frac{\omega_{RF}^{2}}{2\omega_{Z}}\left(1-\text{sinc}(\omega_{Z}T_{p})\right)\widehat{I}_{Z}
\label{eq:Havg1}
\end{eqnarray}
$\widehat{P}^{\text{exact}}_{\phi}(T_{p})$ in Eq. (\ref{eq:pulprop}) can be approximated as:
\begin{eqnarray}
\widehat{P}_{\phi}^{\text{exact}}(T_{p})&\approx&\widehat{P}_{\phi}(T_{p})\approx \exp\left(-iT_{p}\omega_{Z}\widehat{I}_{Z}\right)\exp\left(-\frac{iT_{p}}{\hbar}\left(\overline{H}_{p,\phi}^{(1)}+\overline{H}_{p,\phi}^{(2)}\right)\right)
\label{eq:pulapprox}
\end{eqnarray}
For $\omega_{RF}T_{p}\leq \frac{2\pi}{9}$, $||\widehat{P}_{\phi}^{\text{exact}}(T_{p})-\widehat{P}_{\phi}(T_{p})||\leq 10^{-3}$ for all $\omega_{Z}$, where $||A||=\sqrt{\text{Tr}[A^{\dagger}A]}$ represents the Frobenius matrix norm (if $A$ represents the difference of two unitary matrices, then the maximum value of $||A||$ is
$2\sqrt{n}$ where $n$ is the matrix dimension).  Since the DANTE and p-DANTE sequences both consist of a series of small flip-angle $\theta-$pulses with $\theta<\frac{2\pi}{9}$, the approximation $\widehat{P}^{\text{exact}}_{\phi}(T_{p})\approx \widehat{P}_{\phi}(T_{p})$ in Eq. (\ref{eq:pulapprox}) will be used in the rest of this paper.\\
For future comparison of the propagator in Eq. (\ref{eq:pulprop}) with the propagator for a spin-1/2 evolving under either the DANTE or the p-DANTE sequences in Fig. \ref{fig:figure1}, it is useful to consider an alternative description of $\widehat{P}_{\phi}^{\text{exact}}$ in the interaction frame by  dividing $\widehat{T}\exp\left(-\frac{i}{\hbar}\int^{T_{p}}_{0}\widehat{H}_{\text{INT}}(t')\text{d}t'\right)$ in Eq. (\ref{eq:pulprop}) into $N\gg 1$ smaller propagators, which is illustrated in Figure \ref{fig:figure2}(A).  In this case, $P_{\phi}^{\text{exact}}(T_{p})$ can be rewritten as:
\begin{eqnarray}
\widehat{P}_{\phi}^{\text{exact}}(T_{p})&=&\widehat{U}_{\text{free}}(\omega_{Z}T_{p})\widehat{T}\exp\left(-i\int^{T_{p}}_{0}\text{d}t'\frac{\widehat{H}_{INT}(t')}{\hbar}\right)\nonumber\\
&\approx&\widehat{U}_{\text{free}}(\omega_{Z}T_{p})\widehat{T}\prod_{j=1}^{N}\exp\left(-i\int^{j\frac{T_{p}}{N}}_{(j-1)\frac{T_{p}}{N}}\text{d}t'\frac{\widehat{H}_{INT}(t')}{\hbar}\right)\nonumber\\
&\approx&\widehat{U}_{\text{free}}(\omega_{Z}T_{p})\widehat{T}\prod_{j=1}^{N}\exp\left(-i\frac{T_{p}}{N}\omega_{RF}\text{sinc}\left(\frac{\omega_{Z}T_{p}}{2N}\right)\left(\widehat{I}_{X}\cos\left(\phi_{j}^{*}\right)-I_{Y}\sin\left(\phi_{j}^{*}\right)\right)\right)
\label{eq:puld}
\end{eqnarray}
where $\phi_{j}^{*}=\frac{\omega_{Z}(j-1)T_{p}}{N}+\frac{T_{p}}{2N}-\phi$.  With respect to Eq. (\ref{eq:puld}) and Fig. \ref{fig:figure2}(A), the total propagator for an RF pulse of strength $\omega_{RF}$ applied off-resonance by $\omega_{Z}$ for a time $T_{p}$ is equivalent to the application of $N$ continuous, small-flip $\theta=\frac{T_{p}}{N}\text{sinc}\left(\frac{\omega_{Z}T_{p}}{2N}\right)$, phase-modulated RF pulses where the phase of the $j^{th}$ pulse is $-\phi_{j}^{*}$, followed by a rotation about the $\widehat{z}-$axis by $\omega_{Z}T_{p}$.

%The conditions in which Eq. \ref{eq:pulapprox} is valid are shown in Figure \ref{fig:figure2a}.  A contour plot of $||\widehat{P}_{\phi}^{\text{exact}}-\widehat{P}_{\phi}||$ (where $||A||$ denotes the Frobenius norm given by $||A||=\sqrt{\text{Tr}[A^{\dagger}A]}$) as a function of $\frac{\omega_{Z}}{\omega_{RF}}$ and $\Theta=\omega_{RF}t_{p}$ (the total pulse flip-angle for irradiation on resonance, i.e., $\omega_{Z}=0$) is shown for forty equally spaced contours from $||\widehat{P}^{\text{exact}}_{\phi}-\widehat{P}_{\phi}||=0.0118$ to $||\widehat{P}^{\text{exact}}_{\phi}-\widehat{P}_{\phi}||=0.472$, where the maximum Frobenius norm for the difference of two unitary matrices is $2\sqrt{n}$, where $n$ is the dimension of the matrix.

%In general, using only the first two terms of $\overline{H}_{p,\phi}$ in $\widehat{P}_{\phi}$ is a good approximation to $\widehat{P}_{\phi}^{\text{exact}}$ when $\omega_{Z}\gg \omega_{X}$ (a la the rotating wave approximation).  However,
%when $\omega_{Z}$ is close to zero, $\overline{P}_{\phi}\approx\widehat{P}_{\phi}^{\text{exact}}$ since in this case, $\widehat{H}_{INT}(t)\approx \widehat{H}$.  Figure \ref{fig:figureerr}(A)

%  \begin{figure}%[b!]
%\includegraphics*[height=10cm,width = 10cm]{conpulsep}
%{\small{\caption{ Forty, equally spaced contours from $||\widehat{P}_{\phi}^{\text{exact}}-\widehat{P}_{\phi}||=0.0118$ to $||\widehat{P}_{\phi}^{\text{exact}}-\widehat{P}_{\phi}||=0.472$ \label{fig:figure2a}}}}
%\end{figure}

\subsection{DANTE Pulse Sequence}
\begin{figure}%[b!]
\includegraphics*[height=8cm,width = 12cm]{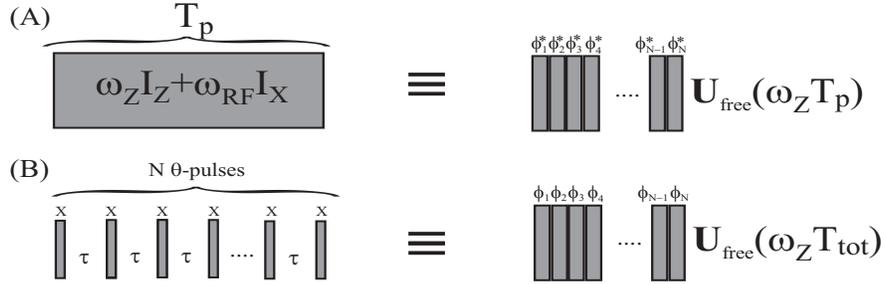}
{\small{\caption{The connection between the DANTE pulse sequence and the application of an off-resonant RF pulse.  (A) An RF pulse applied off resonance by $\omega_{Z}$ and with strength $\omega_{RF}$ for a time $T_{p}$ is equivalent to applying $N\gg 1$ phase-modulated, small flip-angle $\theta=\omega_{RF}\frac{T_{p}}{N}\text{sinc}\left(\frac{\omega_{Z}T_{p}}{2N}\right)$ pulses followed by a rotation about the $\widehat{z}$-axis by an angle $\omega_{Z}T_{p}$.  The phase of the $j^{th}$ pulse is given by $-\phi_{j}^{*}=(j-1)\omega_{Z}\frac{T_{p}}{N}+\omega_{Z}\frac{T_{p}}{2N}-\phi$ (in Fig. \ref{fig:figure2}(A), $\phi=0$).  In (B), a DANTE pulse sequence is equivalent to a series of $N$ phase-modulated
small-flip pulses followed by a rotation about the $\widehat{z}-$axis by an angle $N\omega_{Z}\tau$ [Eq. (\ref{eq:prop})].  The phase
of the $k^{th}$ pulse is $\phi_{k}=\omega_{Z}\left((k-1)(\tau_{t})+\frac{t_{p}}{2}\right)$ where $\tau_{t}=\tau+t_{p}$, $t_{p}$ is the length of the small-flip pulses, and $T_{\text{tot}}=(N-1)\tau_{t}+t_{p}$.  \label{fig:figure2}}}}
\end{figure}

The DANTE sequence\cite{Bodenhausen76} consists of a series of $N$, equally spaced small-tip, $\theta$-pulses of constant phase and duration $t_{p}$ where $\omega_{RF}t_{p}=\theta \ll 1$ [Figure \ref{fig:figure1}(A)].  The full propagator for the DANTE pulse sequence can be written as:
\begin{eqnarray}
\widehat{U}_{\text{exact}}(T_{\text{tot}})&=&\widehat{P}_{0}(t_{p})\left(\widehat{U}_{\text{free}}(\omega_{Z}\tau)\widehat{P}_{0}(t_{p})\right)^{N-1}=\left(\widehat{P}_{0}(t_{p})\widehat{U}_{\text{free}}(\omega_{Z}\tau)\right)^{N-1}\widehat{P}_{0}(t_{p})\nonumber\\
&=&\widehat{U}_{\text{free}}(\omega_{Z}\left((N-1)\tau_{t}+t_{p}\right))\widehat{T}\prod_{k=0}^{N-1}\widehat{P}_{-k\omega_{Z}\tau_{t}}(t_{p})
\label{eq:prop}
\end{eqnarray}
where $\tau$ is the time delay between pulses, $\tau_{t}=\tau+t_{p}$ and \begin{eqnarray}
\widehat{P}_{-k\omega_{Z}\tau_{t}}(t_{p})&=&\widehat{U}^{\dagger}_{\text{free}}(\omega_{Z}k\tau_{t})\widehat{P}_{0}(t_{p})\widehat{U}_{\text{free}}(\omega_{Z}k\tau_{t})\equiv\exp\left(-i\widehat{H}_{k}t_{p}\right)\\
\widehat{H}_{k}&=&\overline{H}_{p,-k\omega_{Z}\tau_{t}}\nonumber\\
&\approx&\frac{\omega_{RF}}{2}\text{sinc}\left(\frac{\omega_{Z}t_{p}}{2}\right)\left(\widehat{I}_{+}e^{i\omega_{Z}\frac{2k\tau_{t}+t_{p}}{2}}+\widehat{I}_{-}e^{-i\omega_{Z}\frac{2k\tau_{t}+t_{p}}{2}}\right)+\frac{\omega_{RF}^{2}}{2\omega_{Z}}\left(1-\text{sinc}(\omega_{Z}t_{p})\right)\widehat{I}_{Z}\nonumber\\
\label{eq:Hk}
\end{eqnarray}
As has been previously noted\cite{Shinnar87,Shinnar89,Shinnar89a}, the propagator for the DANTE sequence in Eq. (\ref{eq:prop}) is the same as the propagator for a continuous series of $N$,
phase modulated small-flip pulses, where the phase modulation depends upon the spin's chemical shift, $\omega_{Z}$, followed by a rotation about the $\widehat{z}-$ axis by $\omega_{Z}T_{\text{tot}}$.  This is illustrated in Figure \ref{fig:figure2}(B).
  If $\text{mod}\left[\omega_{Z}\tau_{t},2\pi\right]\approx 0$,  then all $N$ pulses are effectively
  applied along the same direction since $\phi_{k}\approx \phi_{j}$ for all $k$ and $j$.  The small rotations are therefore additive and lead to an overall rotation of $\Theta\approx N\theta$ about an axis in the transverse plane is generated.  When $\text{mod}\left[\omega_{Z}\tau_{t},2\pi\right]\neq 0$,
 the pulses are effectively applied about different directions ($\phi_{k}\neq \phi_{j}$ for $k\neq j$ in general) thereby reducing the overall spin rotation.
Comparing Fig. \ref{fig:figure2}(B) and Eqs. (\ref{eq:prop}) and (\ref{eq:Hk}) with Fig. \ref{fig:figure2}(A) and Eq. (\ref{eq:puld}), the propagator for the DANTE sequence is similar to that of an off resonant, RF pulse of duration $Nt_{p}=T_{p}$ followed by a rotation about the $\widehat{z}$-axis.  That is,
\begin{eqnarray}
\widehat{U}_{\text{exact}}(T_{\text{tot}})&\approx&\widehat{U}_{\text{free}}(\omega_{Z}T_{\text{tot}}-\omega_{Z}'Nt_{p})\exp\left[-iT_{p}\left(\omega_{Z}'\widehat{I}_{Z}+\omega_{RF}'\left(\widehat{I}_{X}\cos(\phi')+\widehat{I}_{Y}\sin(\phi')\right)\right)\right]\nonumber\\
\end{eqnarray}
where $\omega_{Z}'=\frac{2\pi}{t_{p}}\text{mod}\left[\omega_{Z}\tau_{t},2\pi\right]$, $\omega_{RF}'=\frac{\text{sinc}\left(\frac{\omega_{Z}t_{p}}{2}\right)}{\text{sinc}\left(\frac{\omega_{Z}'t_{p}}{2}\right)}\omega_{RF}$, and $\phi'=\frac{t_{p}}{2}(\omega_{Z}'-\omega_{Z})$.  As mentioned above, when $\text{mod}\left[\omega_{Z}\tau_{t},2\pi\right]=0$, then $\omega_{Z}'=0$, and the effective pulse is applied on resonance and rotates the spin by $\Theta$.  When $\text{mod}\left[\omega_{Z}\tau_{t},2\pi\right]\neq 0$, then $\omega_{Z}'$ can be quite large since $\frac{2\pi}{t_{p}}\gg 1$ for short pulses ($t_{p}\ll 1$).  In this case, the pulse appears to be applied very far off resonance when $\omega_{Z}'\gg \omega_{RF}'$.

%analytical expressions for the propagator under the DANTE pulse sequence in the absence of resonance offset during the pulse exist\cite{Canet95}, in following
%we will use AHT to rewrite the DANTE pulse propagator as To make the above arguments more quantitative, an effective Hamiltonian for the propagator in Eq. (\ref{eq:prop})
%For small flip angle pulses, i.e., $|\widehat{H}_{k}t_{p}|\ll 1$, $\widehat{U}(T_{\text{tot}})$ in Eq. (\ref{eq:prop}) can be approximated as  $\widehat{U}(T_{\text{tot}})=\widehat{U}_{\text{free}}(T_{\text{tot}})\exp\left(-iNt_{p}\overline{H}_{avg}\right)$ where $\overline{H}_{avg}=\sum_{n=1}^{\infty}\overline{H}_{avg}^{(n)}$  with the first three terms of the series given by:

In order to make the above arguments more quantitative, AHT can be used to rewrite $\widehat{U}(T_{\text{tot}})$ in Eq. (\ref{eq:prop}) as $\widehat{U}(T_{\text{tot}})\approx \widehat{U}_{\text{AHT}}(T_{\text{tot}})$ where $\widehat{U}_{\text{AHT}}(T_{\text{tot}})=\widehat{U}_{\text{free}}(\omega_{Z}T_{\text{tot}})\exp\left(-iNt_{p}\overline{H}_{avg}\right)$, where the first two terms in the average Hamiltonian, $\overline{H}_{avg}=\sum_{n=1}^{\infty}\overline{H}_{avg}^{(n)}$, are given by [setting $a=\omega_{RF}\text{sinc}\left(\frac{\omega_{Z}t_{p}}{2}\right)$ and $b=\frac{\omega_{RF}^{2}}{2\omega_{Z}}\left(1-\text{sinc}\left(\omega_{Z}t_{p}\right)\right)$]:
{\footnotesize{\begin{eqnarray}
\overline{H}_{avg}^{(1)}&=&\frac{1}{Nt_{p}}\sum_{k=1}^{N-1}t_{p}\widehat{H}_{k}\nonumber\\
&=&a\frac{\text{sinc}\left(\frac{N\omega_{Z}\tau_{t}}{2}\right)}{\text{sinc}\left(\frac{\omega_{Z}\tau_{t}}{2}\right)}\widehat{I}_{T}(\omega_{Z},\tau_{t},t_{p},N)+b\widehat{I}_{Z}\nonumber\\
\overline{H}_{avg}^{(2)}&=&\frac{1}{2iNt_{p}}\sum_{k>j}\left[\widehat{H}_{k}t_{p},\widehat{H}_{j}t_{p}\right]=-\frac{a^{2}t_{p}}{2\omega_{Z}\tau_{t}}\frac{\text{sinc}\left(N\omega_{Z}\tau_{t}\right)-\text{sinc}(\omega_{Z}\tau_{t})}{\left(\text{sinc}\left(\frac{\omega_{Z}\tau_{t}}{2}\right)\right)^{2}}\widehat{I}_{Z}\nonumber\\
&+&\frac{abt_{p}(N^2-1)}{4N\omega_{Z}\tau_{t}}\frac{\text{sinc}\left(\frac{(N+1)\omega_{Z}\tau_{t}}{2}\right)-\text{sinc}\left(\frac{(N-1)\omega_{Z}\tau_{t}}{2}\right)}{\text{sinc}^{2}\left(\frac{\omega_{Z}\tau_{t}}{2}\right)}\widehat{I}_{T}(\omega_{Z},\tau_{t},t_{p},N)\nonumber\\
\label{eq:Havg}
\end{eqnarray}}}
where $\widehat{I}_{T}(\omega_{Z},\tau_{t},t_{p},N)=\widehat{I}_{X}\cos\left(\frac{\omega_{Z}T_{\text{tot}}}{2}\right)-\widehat{I}_{Y}\sin\left(\frac{\omega_{Z}T_{\text{tot}}}{2}\right)$.
Although the form of $\overline{H}_{avg}$ in Eq. (\ref{eq:Havg}) is somewhat complicated, the physical picture behind $\overline{H}_{avg}$ in Eq. (\ref{eq:Havg}) can be seen in Fig. \ref{fig:figure2}(B).
When $\omega_{Z}\tau_{t}=2\pi n$ for integer $n$, $\phi_{j}=\phi_{k}$ and therefore $\left[\widehat{H}_{k},\widehat{H}_{j}\right]=0$ for all $k\neq j$, so $\overline{H}_{avg}=\overline{H}_{avg}^{(1)}$ exactly.  In this case, the propagator represents a rotation about an axis in the transverse plane of phase $(-1)^n\omega_{Z}t_{p}/2$ by a total angle of
$\Theta=N\omega_{RF}t_{p}\text{sinc}\left(\omega_{Z}t_{p}/2\right)\approx N\omega_{RF}t_{p}=N\theta$ for $\omega_{Z}t_{p}\ll 1$.  For $\omega_{Z}\tau_{t}\neq 2\pi n$ for integer $n$, the various $\widehat{H}_{k}$ are pointing in different directions, so that
the average transverse field in $\overline{H}_{avg}^{(1)}$ is lessened.  Furthermore, since the effective rotation directions no longer commute with one another, i.e., $[\widehat{H}_{k},\widehat{H}_{j}]\neq 0$ for
$k\neq j$, there is a contribution to $\overline{H}_{avg}$ at second-order, $\overline{H}_{avg}^{(2)}$,  of an effective field along the $\widehat{z}$ direction.  When $|\overline{H}_{avg}^{(2)}|\gg |\overline{H}_{avg}^{(1)}|$, the effective field
lies mostly about the $\widehat{z}-$axis and the spins are minimally excited (this argument is similar to the concept of second-averaging\cite{Dybowski75}).

\begin{figure}%[b!]
\includegraphics*[height=9cm,width = 13cm]{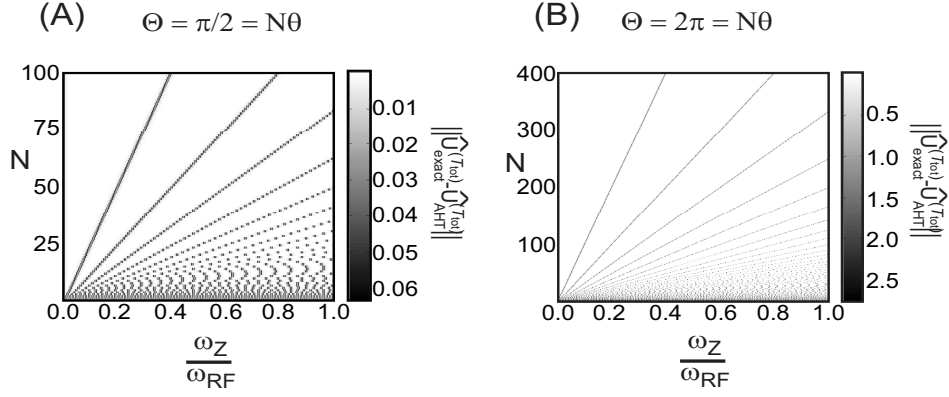}
{\small{\caption{The Frobenius norm of $||\widehat{U}_{\text{exact}}(T_{\text{tot}})-\widehat{U}_{\text{AHT}}(T_{\text{tot}})||$ as a function of the number of small tip $\theta$-pulses, $N$, with $\theta=\frac{\Theta}{N}$, and $\frac{\omega_{Z}}{\omega_{RF}}$.  In Fig. \ref{fig:figure3}, two calculations are shown for total maximum flip angle of (A) $\Theta=\frac{\pi}{2}$ and (B) $\Theta=2\pi$.  $\widehat{U}_{\text{exact}}(T_{\text{tot}})$ [Eq. (\ref{eq:prop})] is the exact propagator for the
DANTE sequence, and $\widehat{U}_{\text{AHT}}(T_{\text{tot}})$ is the propagator calculated using the average Hamiltonian, $\overline{H}_{avg}\approx\overline{H}_{avg}^{(1)}+\overline{H}_{avg}^{(2)}$ in Eq. (\ref{eq:Havg}).
In both calculations, $\frac{\tau}{t_{p}}=1000$. The greatest deviations between $\widehat{U}_{\text{exact}}(T_{\text{tot}})$ and $\widehat{U}_{\text{AHT}}(T_{\text{tot}})$ occur about the resonance conditions in Eq. (\ref{eq:resono}), and the maxima in $||\widehat{U}_{\text{exact}}(T_{\text{tot}})-\widehat{U}_{\text{AHT}}(T_{\text{tot}})||$ are approximately described by two parallel lines given by $N=\text{INT}\left[\frac{\omega_{Z}}{\omega_{RF}}\frac{\Theta}{2\pi n}\left(\frac{\tau}{t_{p}}+1\right)\pm \frac{3}{5n}\right]$ for integer $n\neq 0$.
The agreement between $\widehat{U}_{\text{AHT}}(T_{\text{tot}})$ and $\widehat{U}_{\text{exact}}(T_{\text{tot}})$ improves with decreasing $\Theta$.         \label{fig:figure3}}}}
\end{figure}

In order to see under what conditions AHT can be used in calculating the DANTE pulse sequence, Figure \ref{fig:figure3}
shows $||\widehat{U}_{\text{exact}}(T_{\text{tot}})-\widehat{U}_{\text{AHT}}(T_{\text{tot}})||$ using $\overline{H}_{\text{avg}}\approx \overline{H}_{\text{avg}}^{(1)}+\overline{H}_{\text{avg}}^{(2)}$, as a function of $N$ and $\omega_{Z}/\omega_{RF}$.  Two calculations are shown in Fig. \ref{fig:figure3}, one for a total pulse rotation for an on-resonant RF pulse of $\Theta=\pi/2$ [Fig. \ref{fig:figure3}(A)] and one for $\Theta=2\pi$ [Fig. \ref{fig:figure3}(B)].  For both calculations,
$\frac{\tau}{t_{p}}=1000$. As mentioned above, $\widehat{U}_{\text{AHT}}(T_{\text{tot}})\approx \widehat{U}_{\text{exact}}(T_{\text{tot}})$
when $2\pi n=\omega_{Z}\tau_{t}=\frac{\omega_{Z}}{\omega_{RF}}\omega_{RF}t_{p}\left(\frac{\tau}{t_{p}}+1\right)=\frac{\omega_{Z}}{\omega_{RF}}\frac{\Theta}{N}\left(\frac{\tau}{t_{p}}+1\right)$, where $n$ is an integer.   Therefore, at a resonance condition for $n\neq 0$, there
exists a linear relationship
between $N$ and $\frac{\omega_{Z}}{\omega_{RF}}$ that is given by
\begin{eqnarray}
N&=&\text{INT}\left[\frac{\omega_{Z}}{\omega_{RF}}\frac{\Theta}{2\pi n}\left(\frac{\tau}{t_{p}}+1\right)\right]
\label{eq:resono}
\end{eqnarray}
where $\text{INT}[x]$ gives the nearest integer near $x$.  From Fig. \ref{fig:figure3}, for $\theta\leq \frac{\pi}{60}$ [$N\geq 30$ in Fig. \ref{fig:figure3}(A) and $N\geq 120$ in Fig. \ref{fig:figure3}(B)], $\widehat{U}_{\text{AHT}}(T_{\text{tot}})$ is a good
approximation to $\widehat{U}_{\text{exact}}(T_{\text{tot}})$ for all $\frac{\omega_{Z}}{\omega_{RF}}$ except near the resonance conditions in Eq. (\ref{eq:resono}).  The deviations of $\widehat{U}_{\text{AHT}}(T_{\text{Tot}})$ from $\widehat{U}_{\text{exact}}(T_{\text{tot}})$ occur when $\omega_{Z}$ is slightly
away from the resonance condition, $\omega_{Z}=\frac{2\pi n}{\tau_{t}}$ for integer $n$.  From numerical calculations, the range of frequencies in which
$\widehat{U}_{\text{AHT}}(T_{\text{Tot}})$ is a good approximation to $\widehat{U}_{\text{exact}}(T_{\text{tot}})$
 is found to be approximately given by $\delta\omega_{Z}\gg \frac{6\theta\pi}{5\tau_{t}\Theta}$ or $\delta\omega_{Z}\ll \frac{6\theta\pi}{5\tau_{t}\Theta}$, where
$\delta\omega_{Z}=\text{min}\left[|\omega_{Z}-\frac{2\pi n}{\tau_{t}}|\right]$ is the smallest frequency difference between $\omega_{Z}$ and the nearest integer multiple of the DANTE resonance frequency, $\frac{2\pi}{\tau_{t}}$.  Although the range of $\omega_{Z}$ where $\widehat{U}_{\text{AHT}}(T_{\text{tot}})$ is a good approximation increases with decreasing $\theta$ (increasing $N$), the maximum of $||\widehat{U}_{\text{AHT}}(T_{tot})-\widehat{U}_{\text{exact}}(T_{tot})||$ mainly depends upon the overall rotation angle, $\Theta$.  For $\Theta<\frac{5\pi}{9}$, $\text{max}\left[||\widehat{U}_{\text{AHT}}(T_{tot})-\widehat{U}_{\text{exact}}(T_{tot})||\right]\leq 0.1$ for all $N$ and $\omega_{Z}$.
Physically, this can be understood as follows:
for $\delta\omega_{Z}t_{p}\gg \frac{6\theta\pi}{5\tau_{t}\Theta}$, the effective phases of the pulses [see Fig. \ref{fig:figure2}(B)] are modulated faster than the effective tip of the pulse, $\omega_{RF}t_{p}=\theta$, so that $\omega_{Z}\gg \omega_{RF}$.  In this case,
AHT works well, just as in using the typical rotating wave approximation.  For $\omega_{Z}\ll \frac{6\pi\theta}{5\tau_{t}\Theta}$, the phases of the pulses in Fig. \ref{fig:figure2}(B) are relatively unchanged during the course of the sequence;
 in this case, the various Hamiltonians, $\widehat{H}_{k}$ in Eq. (\ref{eq:Hk}) commute with one another, so $\overline{H}_{avg}\approx \overline{H}_{avg}^{(1)}$ and $\widehat{U}_{\text{exact}}(T_{tot})\approx \widehat{U}_{\text{AHT}}(T_{tot})$.

\subsection{pseudorandom DANTE (p-DANTE)}  In the DANTE sequence, a natural frequency of $\frac{1}{\tau_{t}}$ is
introduced into the dynamics due to the periodicity of the pulse sequence;  this leads to efficient excitation at
 frequencies $\nu^{\text{DANTE}}_{n}=\frac{n}{\tau_{t}}$ for integer $n$.  However, suppose that one was interested in
 using a DANTE-like sequence to efficiently excite only one particular frequency, say at $\nu^{\text{DANTE}}_{0}=0$ Hz.
 One way to accomplish this using a DANTE sequence would be to make $\tau_{t}$ small enough such that all $\nu^{\text{DANTE}}_{n\neq 0}$ lie
 outside the relevant spectral width.  If the spectral width for the system of interest is large, however, this would necessitate using small $\tau_{t}$, where the smallest time possible $\tau_{t}$ is $\tau_t\approx t_{p}$ (i.e., when $\tau=0$).    The selectivity or width of the excitation spectrum about $\nu^{\text{DANTE}}_{n}$ is approximately given by $\frac{1}{N\tau_{t}}$.  For $\tau_{t}\approx t_{p}$, this means the selectivity is roughly proportional to $\frac{1}{Nt_{p}}=\frac{2\pi\nu_{RF}}{\Theta}$.  In this limit,
 the effect of the DANTE sequence is similar to evolution under continuous RF irradiation, which leads to a very broad excitation profile unless $\nu_{RF}$ is weak or
 $\Theta\gg 2\pi$.  Under these conditions, the DANTE sequence would be equivalent to applying a long, low-amplitude RF pulse.

    An alternative way to excite only a single resonance using a DANTE-like sequence would be to violate the periodicity of the DANTE sequence.  This could be accomplished in a variety of ways, such as using aperiodic delays, modulating the pulse amplitudes and delays, etc.  One example of such an aperiodic DANTE sequence is illustrated in Figure \ref{fig:figure1}(B) where $N$ small-flip $\theta$-pulses are applied with a delay of $\tau_{k}$ between the separation between the $k^{th}$ and $(k+1)^{th}$ pulse, where in general, $\tau_{k}\neq \tau_{j}$.  For such pulse sequences to selectively and efficiently excite spins at a single resonance frequency $\nu_{0}$, the phases of the pulses, $\phi_{k}$, must be modulated.  In this case, the phase of the $(k+1)^{th}$ pulse is given by  $\phi_{k+1}=-2\pi\nu_{0}\left(kt_{p}+\sum_{j=1}^{k}\tau_{j}\right)=-2\pi\nu_{0}T_{k}$ with $T_{k}=kt_{p}+\sum_{j=1}^{k}\tau_{j}$, and $\phi_{1}=0$ and $T_{0}=0$.  Such a set of aperiodic DANTE sequences are referred to as pseudorandom-DANTE or p-DANTE sequences.

\begin{figure}%[b!]
\includegraphics*[height=10cm,width = 10cm]{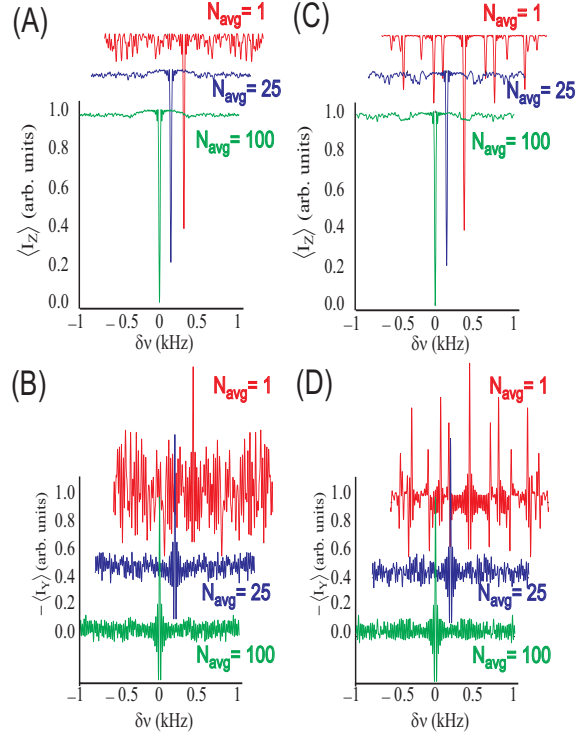}
{\small{\caption{Theoretically calculated "averaged" excitation [(B) and (D)] and $\widehat{z}-$magnetization profiles [(A) and (C)] for a series of p-DANTE sequences [Fig. \ref{fig:figure1}(B)].  In all p-DANTE sequences, $N=30$, $\theta=\frac{\pi}{60}$, and $\phi_{k}=0$ for all $k$ so that only spins resonating at $\delta \nu =0$ Hz are rotated about the $\widehat{x}$-axis by $\Theta=\frac{\pi}{2}$.  In (A) and (B), one hundred different p-DANTE sequences with randomly chosen delays were generated such that the average pulse delay, $\frac{1}{29}\sum_{k=1}^{29}\tau_{k}=1.6$ ms, was the same for all p-DANTE sequences.  The excitation and $\widehat{z}$-magnetization profiles were averaged over $N_{avg}$ p-DANTE sequences with $N_{avg}=1$ (red curve), $N_{avg}=25$ (blue curve), and $N_{avg}=100$ (green curve(. In (C) and (D), one hundred different p-DANTE sequences were chosen such that the $k^{th}$ delay for the $p^{th}$ experiment was $\tau^{p}_{k}=\tau_{p}\left[1+\frac{\delta\tau}{\tau}\cos\left(\frac{2\pi k}{f_{p}}\right)\right]$ where $\frac{\delta\tau}{\tau}=\frac{1}{\sqrt{2}}$ and
$f_{p}$ was inversely proportional to the square root of the $p^{th}$ prime number (e.g., $f_{1}=1/\sqrt{2}$, $f_{2}=1/\sqrt{3}$, $f_{100}=1/\sqrt{541}$).  $\tau_{p}$ was chosen to ensure that the averaged delay was $\frac{1}{29}\sum_{k=1}^{29}\tau^{p}_{k}=1.6$ ms for all $p$.  The averaged excitation and $\widehat{z}$-magnetization profiles for $N_{avg}=1$ (red curve with $f_{1}=1/\sqrt{2}$), $N_{avg}=25$ (blue curve, averaging from $f_{1}=1/\sqrt{2}$ to $f_{25}=1/\sqrt{97}$), and $N_{avg}=100$ (green curve, averaging from $f_{1}=1/\sqrt{2}$ to $f_{100}=1/sqrt{541}$) are shown.  As $N_{avg}$ increases, the "fluctuations" in both $\langle I_{Z}\rangle$ and $-\langle I_{Y}\rangle$ decrease, and the results are similar for p-DANTE sequences using randomly chosen delays [(A) and (B)] and those using periodically modulated delays [(C) and (D)].  \label{fig:figure4}}}}
\end{figure}

      As in the DANTE case, the propagator for the p-DANTE sequence can be written as $\widehat{U}(T_{\text{tot}})=\widehat{U}_{\text{free}}(\omega_{Z}(T_{\text{tot}}))\exp\left(-\frac{i}{\hbar}Nt_{p}\overline{H}_{avg}\right)$, where $T_{\text{tot}}=T_{N-1}+t_{p}$, and $\overline{H}_{avg}$ is the average Hamiltonian for the p-DANTE sequence, with the first two terms given by:
  {\footnotesize{\begin{eqnarray}
  \label{eq:prandH1}
    \overline{H}^{(1)}_{\text{avg}}&=&\frac{a}{N}\sum_{k=1}^{N}\left(\widehat{I}_{+}e^{i\left(\Delta\omega T_{k}+\frac{\omega_{Z}t_{p}}{2}\right)}+I_{-}e^{-i\left(\Delta\omega T_{k}+\frac{\omega_{Z}t_{p}}{2}\right)}\right)+b\widehat{I}_{Z}\\
 \overline{H}^{(2)}_{\text{avg}}&=&\frac{abt_{p}}{2N}\left[\sum_{j<k}\sin\left(\frac{\Delta\omega(T_{j}-T_{k})}{2}\right)\left(\widehat{I}_{+}e^{i\frac{\Delta\omega(T_{j}+T_{k})+\omega_{Z}t_{p}}{2}}+\widehat{I}_{-}e^{-i\frac{\Delta\omega(T_{j}+T_{k})+\omega_{Z}t_{p}}{2}}\right)\right]\nonumber\\
&+&\frac{a^{2}t_{p}}{2N}\widehat{I}_{Z}\sum_{j<k}\sin\left(\Delta\omega(T_{k}-T_{j})\right)
 %&+&\frac{\omega_{RF}^3}{4N\omega_{Z}}\text{sinc}\left(\frac{\omega_{Z}t_{p}}{2}\right)(1-\text{sinc}(\omega_{Z}t_{p}))\left[\widehat{I}_{+}e^{i\frac{\omega_{Z}t_{p}}{2}}\sum_{j<k}e^{i\frac{\Delta\omega(T_{j}+T_{k})}{2}}\sin\left(\frac{\Delta\omega(T_{j}-T_{k})}{2}\right)\right]\nonumber\\
 \label{eq:prandH2}
        \end{eqnarray}}}
        where $\Delta\omega=2\pi\left(\nu_{Z}-\nu_{0}\right)$, and $a$ and $b$ were previously defined before Eq. (\ref{eq:Havg}).  Unlike in the DANTE case, the average delay between pulses fluctuates within the p-DANTE sequence, i.e., $\frac{T_{k}}{k}\neq \frac{T_{j}}{j}$ for $k\neq j$.  For $\nu_{Z}\neq\nu_{0}$, a p-DANTE sequence is effectively equivalent to applying an RF field with a fluctuating offset [Fig. \ref{fig:figure2}(A)], the result of which is a seemingly random excitation profile for those spins with $\Delta\omega_{Z}\neq 0$.  Spins with $\Delta\omega_{Z}=0$ are rotated by $N\theta=\Theta$.  This is illustrated in Figures \ref{fig:figure4}(A) and \ref{fig:figure4}(B), which show the excitation and $\widehat{z}-$magnetization profiles under a p-DANTE sequence respectively.  In Figs. \ref{fig:figure4}(A) and \ref{fig:figure4}(B), the various $N-1$ $\tau_{k}$'s were chosen randomly but were scaled to ensure that $\sum_{k=1}^{N-1}\tau_{k}=46.4$ ms, and one hundred different sets of randomly generated p-DANTE sequences were generated. Consider the excitation and $\widehat{z}-$magnetization profile for a single p-DANTE sequence [red curve ($N_{avg}=1$)] shown in Figs. \ref{fig:figure4}(A) and \ref{fig:figure4}(B).  A maximum rotation by $\Theta=\frac{\pi}{2}$ occurs at $\delta\nu=0$, where $-\langle\widehat{I}_{Y}\rangle=1$ and $\langle \widehat{I}_{Z}\rangle=0$.  Away from $\delta\nu=0$, the excitation and $\widehat{z}-$magnetization profiles are quite noisy, but  $|\langle \widehat{I}_{Y}\rangle|<1$ and $\langle \widehat{I}_{Z}\rangle >0$.  Note that the $\widehat{z}-$magnetization profile is less noisy, since rotations away from the $\widehat{z}$ direction go as $\cos(\theta)\approx 1-\frac{\theta^{2}}{2}$ whereas excitations go as $\sin(\theta)\approx \theta$ for $\theta\ll 1$.

         If the excitation and $\widehat{z}-$magnetization profiles are averaged over different p-DANTE sequences that possess the same total pulse length and are all selective for $\nu_{0}$,  then the fluctuations in both the excitation [Fig. \ref{fig:figure4}(B)] and $\widehat{z}-$magnetization [Fig. \ref{fig:figure4}(B)] profiles for $\delta\nu\neq 0$ decrease relative to the excitation at $\delta\nu=0$ roughly as $\frac{1}{\sqrt{N_{avg}}}$.  However, even when $N_{avg}\gg 1$, there still exists a ''baseline'' excitation at $\delta\nu\neq 0$ which is nonzero (averaging simply decreases the fluctuations about the baseline excitation).  The average ''baseline'' excitation is approximately given by $\overline{|\langle\widehat{I}_{Y}\rangle|}\approx \frac{\Theta}{N+2}$ and $\langle \overline{\widehat{I}_{Z}\rangle}=1-\frac{(N+1)\theta^{2}}{2}$.  Thus decreasing $\theta$ and $N$ will decrease the ''baseline'' excitation as $\theta\propto \frac{1}{N}$.  Similar schemes averaging over random sequences have been previously used for stochastic dipolar recoupling\cite{Tycko07,Tycko08}.

    %In the following, we will consider two cases of ''pseudorandom"  DANTE, the case of periodically modulated delays and the case of purely random delays.

Besides randomly chosen delays, averaging over different sets of delays that are periodically modulated can also lead to selective excitation.  Consider a series of delays where the $k^{th}$ delay is given by
     $\tau_{k}=\tau+\delta\tau\cos\left(\frac{2\pi k}{f}\right)$,
     where $f$ is a real number, and $\tau\geq\delta\tau$ so
     that $\tau_{k}\geq 0$.   For such a sequence to selectively excite spins resonating at $\nu_{0}$,
     the phase of the $k^{th}$ pulse must be given by $\phi_{k}=-2\pi\nu_{0}T_{k-1}$ with $T_{0}=0$ and :
    \begin{eqnarray}
    T_{k}&=&\sum_{j=1}^{k}\tau_{k}=k\tau-\frac{\delta\tau}{2}\left(1-\csc\left(\frac{\pi}{f}\right)\sin\left(\frac{2k+1}{f}\pi\right)\right)
    \label{eq:pdTk}
    \end{eqnarray}
    where the total time of the sequence is given by $T_{\text{tot}}=Nt_{p}+T_{N-1}=Nt_{p}+(N-1)\tau-\frac{\delta\tau}{2}+\frac{\delta \tau}{2}\csc\left(\frac{\pi}{f}\right)\sin\left(\frac{\pi(2N-1)}{f}\right)$.\\
    Using the values of $T_{k}$ in Eq. (\ref{eq:pdTk}),
    $\overline{H}_{avg}^{(1)}$ in Eq. (\ref{eq:prandH1}) can be evaluated and is given by:
 {\footnotesize{   \begin{eqnarray}
    \overline{H}^{(1)}_{avg}&=&\frac{a}{2N}\sum_{k=1}^{N}\left(\widehat{I}_{+}e^{i\left(\Delta\omega T_{k}+\frac{\omega_{Z}t_{p}}{2}\right)}+I_{-}e^{-i\left(\Delta\omega T_{k}+\frac{\omega_{Z}t_{p}}{2}\right)}\right)+b\widehat{I}_{Z}\nonumber\\
&=&\frac{a}{2}\sum_{n=-\infty}^{\infty}J_{n}\left(\frac{\Delta\omega\delta\tau}{2}\csc\left(\frac{\pi}{f}\right)\right)\frac{\text{sinc}\left(\frac{N(\Delta\omega\tau+\frac{2n\pi}{f})}{2}\right)}{\text{sinc}\left(\frac{\Delta\omega\tau+\frac{2n\pi}{f}}{2}\right)}\left(I_{+}e^{i\chi_{n}}+I_{-}e^{-i\chi_{n}}\right)+b\widehat{I}_{Z}
 %\overline{H}^{(2)}_{\text{avg}}&=&\frac{\omega_{RF}^{2}t_{p}}{2N}\text{sinc}^{2}\left(\frac{\omega_{Z}t_{p}}{2}\right)\widehat{I}_{Z}\sum_{j<k}\sin\left(\Delta\omega(T_{k}-T_{j})\right)\nonumber\\
 %&+&\frac{\omega_{RF}^3t_{p}}{4N\omega_{Z}}\text{sinc}\left(\frac{\omega_{Z}t_{p}}{2}\right)(1-\text{sinc}(\omega_{Z}t_{p}))\left[\sum_{j<k}\sin\left(\frac{\Delta\omega(T_{j}-T_{k})}{2}\right)\left(\widehat{I}_{+}e^{i\frac{\Delta\omega(T_{j}+T_{k})+\omega_{Z}t_{p}}{2}}+\widehat{I}_{-}e^{-i\frac{\Delta\omega(T_{j}+T_{k})+\omega_{Z}t_{p}}{2}}\right)\right]\nonumber\\
 %%&+&\frac{\omega_{RF}^3}{4N\omega_{Z}}\text{sinc}\left(\frac{\omega_{Z}t_{p}}{2}\right)(1-\text{sinc}(\omega_{Z}t_{p}))\left[\widehat{I}_{+}e^{i\frac{\omega_{Z}t_{p}}{2}}\sum_{j<k}e^{i\frac{\Delta\omega(T_{j}+T_{k})}{2}}\sin\left(\frac{\Delta\omega(T_{j}-T_{k})}{2}\right)\right]\nonumber\\
 \label{eq:prandH}
        \end{eqnarray}}}
    where $\Delta\omega=\omega_{Z}-\omega_{0}$, $J_{n}$ is a bessel function of order $n$, and $\chi_{n}=\Delta\omega\frac{(N-1)\tau-\delta\tau}{2}+\frac{Nn\pi}{f}+\frac{\omega_{Z}t_{p}}{2}$. From Eq. (\ref{eq:prandH}), $\overline{H}^{(1)}_{\text{avg}}$ is maximal at the conditions $2\pi\Delta\nu\tau+\frac{2n\pi}{f}=2m\pi$ or at $\Delta\nu=\frac{m}{\tau}-\frac{n}{f\tau}$ where $m$ and $n$ are integers.  These define the resonance conditions for this type of p-DANTE sequence.  However, $\overline{H}_{avg}^{(1)}$ is scaled by $J_{n}\left(\left(m-\frac{n}{f}\right)\frac{\pi\delta\tau}{\tau}\csc\left(\frac{\pi}{f}\right)\right)$, which is greatest when $m=n=0$.  For $m\neq 0$ and $n\neq 0$, this scaling is less than one, which results in a smaller total rotation.
%water 0.5054 acetone 1, 0.8447 dmso

Figure \ref{fig:figure4}(C) and (D) show the numerically averaged $\widehat{z}$-magnetization and excitation profiles respectively, averaged for up to one hundred different p-DANTE sequences using periodically modulated delays.  In the simulations, $N=30$ and $\frac{\delta\tau}{\tau}=\frac{1}{\sqrt{2}}$.  For the $p^{th}$ p-DANTE sequence, $f_{p}$ was set to be equal to the inverse of the square root of the $p^{th}$ prime number, i.e., $f_{1}=1/\sqrt{2}$, $f_{2}=1/\sqrt{3}$, $f_{100}=1/\sqrt{541}$.  In order to better compare these results to the results for the p-DANTE sequences using random delays [Figs. \ref{fig:figure4}(A) and \ref{fig:figure4}(B)], $\tau_{p}$ for the $p^{th}$ experiment was chosen so that $\frac{1}{29}\sum_{j=1}^{29}\tau_{j}^{p}=1.6$ ms.  First consider the $N_{avg}=1$ case (red curve) in which $f=\frac{1}{\sqrt{2}}$ and $\frac{1}{\tau}=625.13$ Hz.  Unlike the case of using random delays [red curves in Fig. \ref{fig:figure4}(A) and \ref{fig:figure4}(B)] where the resulting excitations appear randomly distributed throughout the spectral range, the excitation profile using periodically modulated delays occur at discrete $\delta\nu$ given by the resonance condition $\delta\nu=625.13\left(m-\sqrt{2}n\right)$ Hz [Eq. (\ref{eq:prandH})].  Note that while the resonance at $\delta\nu=0$ ($m=0$ and $n=0$) is maximally excited ($\langle \widehat{I}_{Z}\rangle=0$ and $-\langle \widehat{I}_{Y}\rangle=1)$, the degree of excitation at other resonance conditions is less.  In particular, the resonances at $\delta\nu=\pm 625.13$ Hz ($m=\pm 1$, $n=0$) are not observed in the calculated profile, since at these conditions, $\overline{H}_{avg}^{(1)}$ is scaled by $J_{0}(-2.3046)=0.053$, whereas the resonances at $\delta\nu=\pm 366.2$ Hz ($n=\mp 1$ and $m=\pm 2$) are clearly observed $(|J_{1}(1.35)|=0.5325$).  As was the case for p-DANTE sequences using randomly chosen delays, averaging over different sets of periodically modulated p-DANTE sequences reduces the excitation for all resonances except at $\delta\nu=0$, which is a common resonance for all p-DANTE sequences.  From Fig. \ref{fig:figure4}, the $\widehat{z}$-magnetization and excitation profiles using periodically modulated delays [Figs. \ref{fig:figure4}(C) and \ref{fig:figure4}(D)] become similar to those using the randomly chosen delays [Figs. \ref{fig:figure4}(A) and \ref{fig:figure4}(D)] as $N_{avg}$ increases.

\begin{figure}%[b!]
\includegraphics*[height=15cm,width = 8cm]{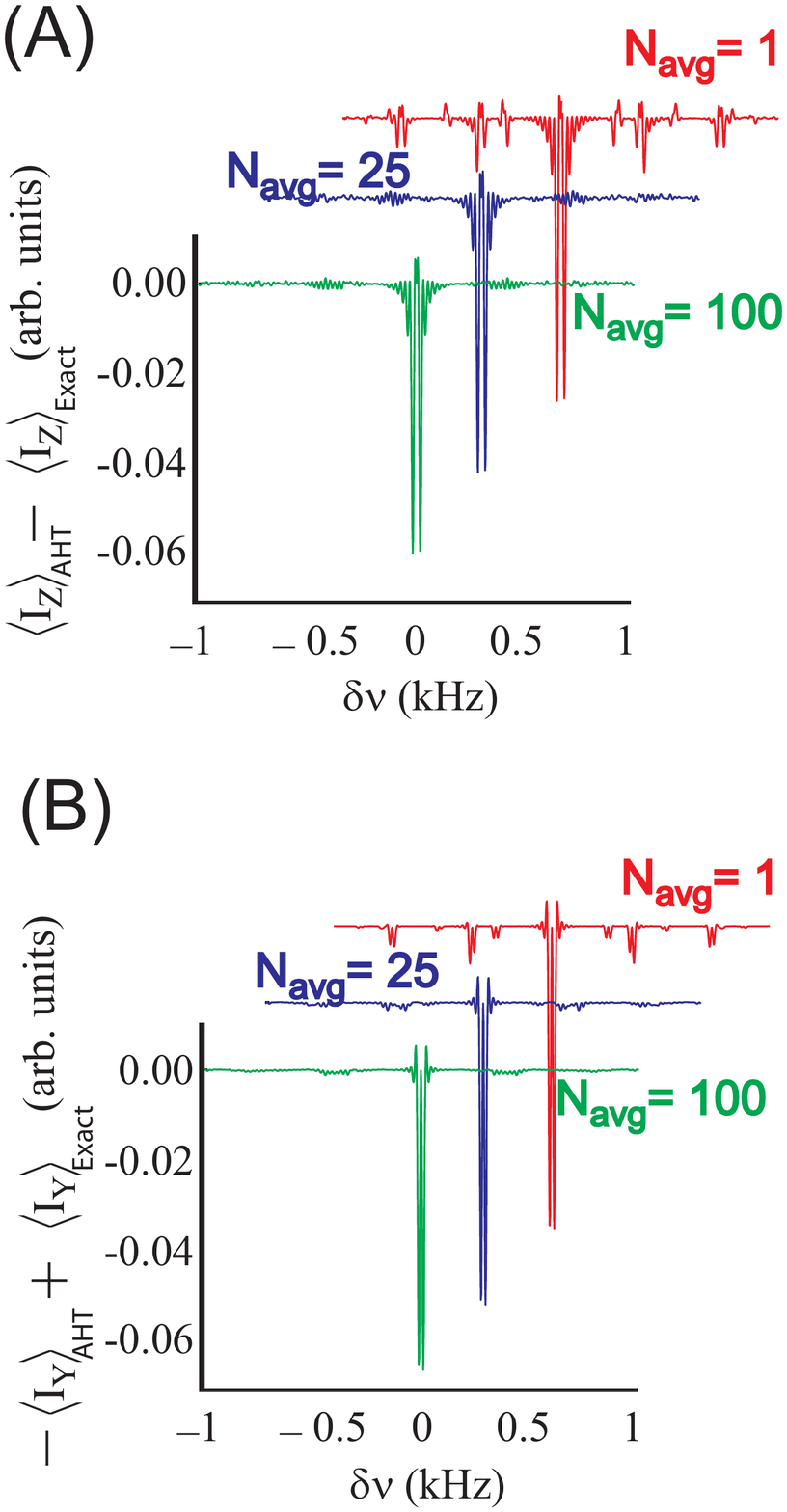}
{\small{\caption{Difference in the calculated (A) $\widehat{z}$-magnetization and (B) excitation profiles for the p-DANTE sequences used in Fig. \ref{fig:figure4}(C) and \ref{fig:figure4}(D) between the profiles calculated using the propagator from AHT, $\widehat{U}_{\text{AHT}}(T_{\text{tot}})=e^{-iT_{\text{Tot}}\omega_{Z}\widehat{I}_{Z}}e^{-iNt_{p}\left(\overline{H}^{(1)}_{\text{avg}}+\overline{H}^{(2)}_{\text{avg}}\right)}$ [Eq. (\ref{eq:prandH}) and Eq. (\ref{eq:prandH2})] and the profiles calculated using exact propagator, $\widehat{U}_{\text{exact}}(T_{\text{tot}})$.  The agreement is relatively good over a wide frequency range except near the $\delta\nu=0$ Hz resonance condition, $\delta\nu\approx \pm\frac{3\theta}{5\Theta\tau_{t}}=\pm 13$ Hz.  For all $|\delta\nu|>>\frac{3\theta}{5\Theta\tau_{t}}$ and $|\delta\nu|\ll \frac{3\theta}{5\Theta\tau_{t}}$, the agreement between AHT and the exact calculation improves with averaging over different p-DANTE sequences.\label{fig:figsim4}}}}
\end{figure}

  Finally, it should be noted that the conditions under which the average Hamiltonian in Eq. (\ref{eq:prandH}) provides a valid description ofthe p-DANTE sequence are approximately the same as those found for the DANTE sequence [Fig. \ref{fig:figure3}].  Figure \ref{fig:figsim4} shows the difference in the excitation and $\widehat{z}$-magnetization profiles calculated using either the exact propagator or the propagator calculated using the average Hamiltonian up to second-order [Eq. \ref{eq:prandH2} and Eq. (\ref{eq:prandH})] for the p-DANTE sequences used in Figs. \ref{fig:figure4}(C) and \ref{fig:figure4}(D). AHT works well for all $\delta\nu$ away from resonance conditions which is evident from Fig. \ref{fig:figsim4} for the $N_{avg}=1$ curve.  The magnitude of the error in this approximation is the same as that found for a DANTE sequence with $\Theta=\pi/2$. The agreement of the AHT calculations with the exact calculations appears to improve upon averaging over different p-DANTE sequences, except for the $\delta\nu=0$ resonance.  This is due to the fact that only the $\delta\nu=0$ resonance is the same for all p-DANTE sequence used in Fig. \ref{fig:figsim4}.

\section{Experimental}  All experiments were performed
on a 300 MHz Avance Bruker spectrometer (static magnetic field of 7 T and an operating frequency for $^{1}H$
of 300.13 MHz), using a 5-mm Bruker BBO probe.  A 2M solution of acetone in acetone$-d_{6}$ was used to experimentally
determine the excitation and $\widehat{z}$-magnetization profiles as a function of frequency offset from the acetone resonance for both the DANTE and two pseudorandom pulse sequences.  The carrier frequency was incremented between $-580$ Hz below to $580$ Hz above the acetone resonance in intervals of
10 Hz in order to experimentally determine the excitation and $\widehat{z}$-magnetization profiles (a total of 1161 measurements), and the integral of the acetone peak was measured.  In order to measure the $\widehat{z}$-magnetization, a $\frac{\pi}{2}$ pulse (Rabi frequency of 21.4 kHz) was applied after the
DANTE and p-DANTE sequences, which was phase cycled in concert with the receiver phase so that only the $\widehat{z}-$magnetization prior to the last $\frac{\pi}{2}$ pulse was measured.  A delay of 40 seconds was used between scans in all experiments in order to ensure that the system had relaxed back to equilibrium which was necessary to avoid any distortions in the observed profiles.

In order to demonstrate the improved selectivity in the excitation and $\widehat{z}-$magnetization profiles by signal averaging over different p-DANTE sequences (as shown in Fig. \ref{fig:figure4}), experiments using different p-DANTE sequences were performed on a solution of
acetone, dimethyl sulfoxide (DMSO),and water diluted in $D_{2}O$, such that $\frac{[\text{Acetone}]}{[\text{DMSO}]}=0.822$ and $\frac{[\text{H}_{2}\text{O}]}{[\text{DMSO}]}=1.4355$.  All chemicals were obtained from Sigma-Aldrich.

\section{Results and Discussion}
\begin{figure}%[b!]
\includegraphics*[height=10cm,width = 7cm]{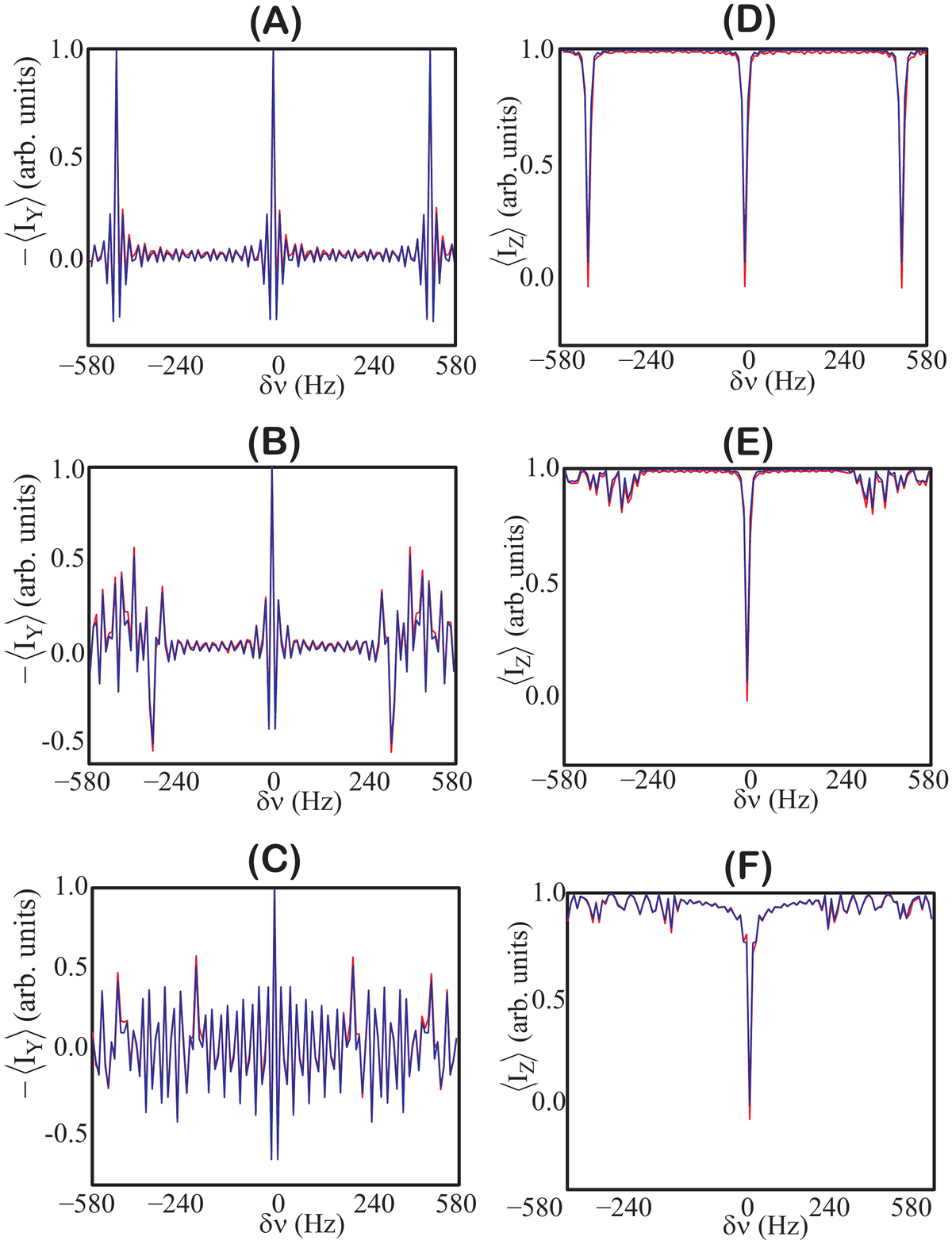}
{\small{\caption{The experimental (red) and theoretical (blue) excitation [(A)-(C)] and $\widehat{z}$-magnetization [(D)-(F)] profiles under application of the DANTE [(A),(D)] and p-DANTE sequences [(B),(E),(C),(F)] as a function of the applied RF's offset away from the acetone resonance, $\delta\nu$.
In all experiments, $N=30$ and $\theta=\frac{\pi}{60}$.  The profiles were generated using a 2M acetone solution in acetone-d$_{6}$ by changing the RF carrier frequency from $-580$ Hz to $580$ Hz in 10 Hz increments, and the resulting acetone resonance was integrated.  For the DANTE sequence, $\tau=2$ ms;  with these parameters, the acetone resonance was maximally excited at $\delta\nu\approx\pm \frac{1}{\tau}=\pm 500$ Hz and at $\delta\nu=0$ Hz over the spectral range $[-580$ Hz, $580$ Hz].  For the p-DANTE sequences, the $k^{th}$ delay was either given by $\tau_{k}=2.096\left[1+\frac{1}{3}\cos\left(\frac{2k\pi}{23}\right)\right]$ms [Figs. \ref{fig:figure6}(B) and \ref{fig:figure6}(E)] or by the UDD\cite{Uhrig09}-like delay $\tau_{k}=2.063\left[1-\cos\left(\frac{k\pi}{29+1}\right)\right]$ms=$4.126\sin^{2}\left(\frac{k\pi}{2(29+1)}\right)$ [Figs. \ref{fig:figure6}(C) and \ref{fig:figure6}(F)].  In both cases, the average delay between pulses was $\frac{1}{29}\sum_{k=1}^{29}\tau_{k}=2$ ms in order to enable comparison with the results from the DANTE sequence in Figs. \ref{fig:figure6}(A) and \ref{fig:figure6}(D).  For both p-DANTE sequences, there is minimal excitation at $\delta\nu=\pm 500$ Hz, although the UDD-like sequence [Figs. \ref{fig:figure6}(C) and \ref{fig:figure6}(F)] appears to generate smaller excitations over a wider frequency range than the other p-DANTE sequence [Figs. \ref{fig:figure6}(B) and \ref{fig:figure6}(E)].  In all cases, there is good agreement between theory (blue) and experiment (red).  \label{fig:figure6}}}}
\end{figure}

 The experimentally determined excitation and $\widehat{z}-$magnetization profiles under the DANTE and two different p-DANTE sequences obtained using a 2M acetone solution in acetone-$d_{6}$ are shown in Figure \ref{fig:figure6}, where the blue and red curves correspond to the theoretical and experimentally observed profiles respectively.  In these experiments, $N=30$, $\theta=\frac{\pi}{60}$, and $t_{p}=720$ ns were used with a maximum rotation of $\Theta=N\theta=\frac{\pi}{2}$.  For the DANTE sequence, $\tau=2$ ms.  Over the spectral range shown in Fig. \ref{fig:figure6}(A) and \ref{fig:figure6}(D), excitations at frequencies $\delta\nu=\frac{\pm 1}{\tau}=\pm 500$ Hz and at $\delta\nu=0$ Hz were observed.  For the p-DANTE sequences, the $k^{th}$ delay was given by either [Figs. \ref{fig:figure6}(C) and \ref{fig:figure6}(F)] $\tau_{k}=2.063\left[1-\cos\left(\frac{k\pi}{29+1}\right)\right]$ms=$4.126\sin^{2}\left(\frac{k\pi}{2(29+1)}\right)$ (which is a similar set of delays used in the UDD sequences\cite{Uhrig09}) or [Figs. \ref{fig:figure6}(B) and \ref{fig:figure6}(E)] $\tau_{k}=2.096\left[1+\frac{1}{3}\cos\left(\frac{2k\pi}{23}\right)\right]$ms.  In both cases, the average delay between pulses was equal to $\frac{1}{29}\sum_{k=1}^{29}\tau_{k}=2$ ms in order to allow for better comparison with the DANTE sequence used in Figs. \ref{fig:figure6}(A) and \ref{fig:figure6}(D).  Both p-DANTE sequences generated a maximum excitation at $\delta\nu=0$ Hz, and smaller excitations for $\delta\nu\neq 0$ Hz were also observed, as expected.  Note that for the UDD-like p-DANTE sequence [Fig. \ref{fig:figure6}(C) and \ref{fig:figure6}(F)], the excitation and $\widehat{z}$-magnetization profiles look similar to that of a p-DANTE sequence using randomly chosen delays [the red curve in Figs. \ref{fig:figure4}(A) and \ref{fig:figure4}(B)], whereas excitations using the other p-DANTE sequence [Figs. \ref{fig:figure6}(B) and \ref{fig:figure6}(E)] appear to be concentrated within a smaller frequency range.\\

\begin{figure}%[b!]
\includegraphics*[height=10cm,width = 6cm]{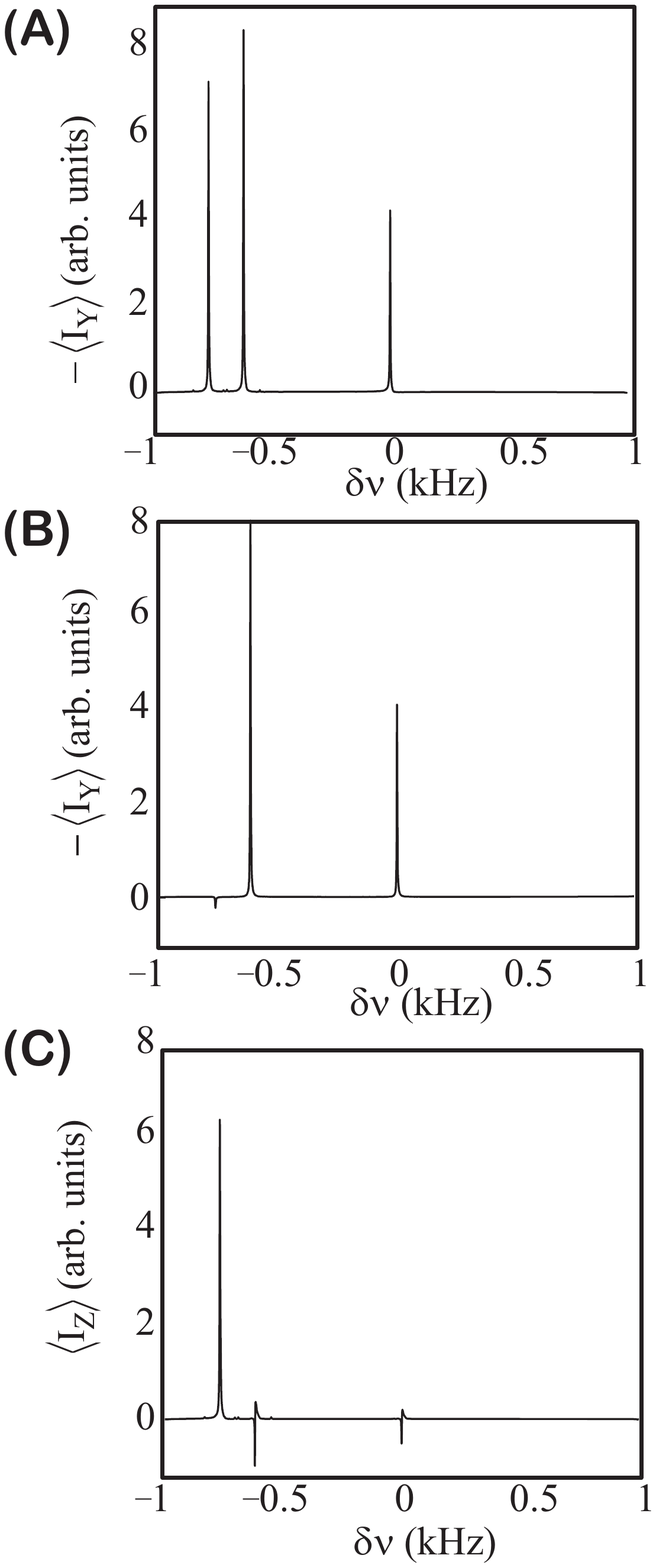}
{\small{\caption{(A) The spectrum after a $\frac{\pi}{2}$-acquire experiment for a DMSO, acetone, and water solution.  The spectrum is centered on the water resonance $(\delta\nu=0)$.  (B) The spectrum after application of a DANTE pulse sequence with $N=30$, $\theta=\frac{\pi}{60}$, $\Theta=N\theta=\frac{\pi}{2}$, and $\tau=\frac{1}{|\delta\nu_{\text{DMSO,water}}|}\approx 1.6$ ms.  In this case, both the water and DMSO resonances are excited whereas very little excitation occurs at the acetone resonance ($\delta\nu_{\text{acetone,water}}=-768$ Hz). (C) The $\widehat{z}-$magnetization weighted spectrum after application of the DANTE pulse sequence.  As expected from (B), there is substantial $\widehat{z}$-magnetization for the acetone resonance and little $\widehat{z}$-magnetization for both the water and DMSO resonances after application of the DANTE sequence.\label{fig:figure7}}}}
\end{figure}

 In order to examine the effects of averaging over different p-DANTE sequences, experiments were performed on a DMSO-acetone-water solution in $D_{2}O$.  The spectrum of the solution after a simple $\frac{\pi}{2}$-acquire sequence is shown in Fig. \ref{fig:figure7}(A), where the RF was applied on resonance with respect to the water resonance [$\delta\nu_{\text{acetone,water}}=-768$ Hz and $\delta\nu_{\text{DMSO,water}}=-620.2$ Hz].  The experimental excitation and $\widehat{z}-$magnetization weighted spectra after the application of a DANTE sequence with $N=30$, $\theta=\frac{\pi}{60}$ ($t_{p}=630$ ns), and $\tau\approx\frac{1}{|\delta\nu_{\text{DMSO,water}}|}=1.6$ ms are shown in Figure \ref{fig:figure7}(B) and Figure \ref{fig:figure7}(C) respectively.  With this choice of $\tau$, the DANTE sequence efficiently excites both the water $\left(\delta\nu=\frac{0}{\tau}\right)$ and DMSO $\left(\delta\nu=-\frac{1}{\tau}\right)$ resonances [Fig. \ref{fig:figure7}(B)] and leaves the acetone magnetization mostly about the $\widehat{z}-$axis [Fig. \ref{fig:figure7}(C)].

\begin{figure}%[b!]
\includegraphics*[height=8cm,width = 14cm]{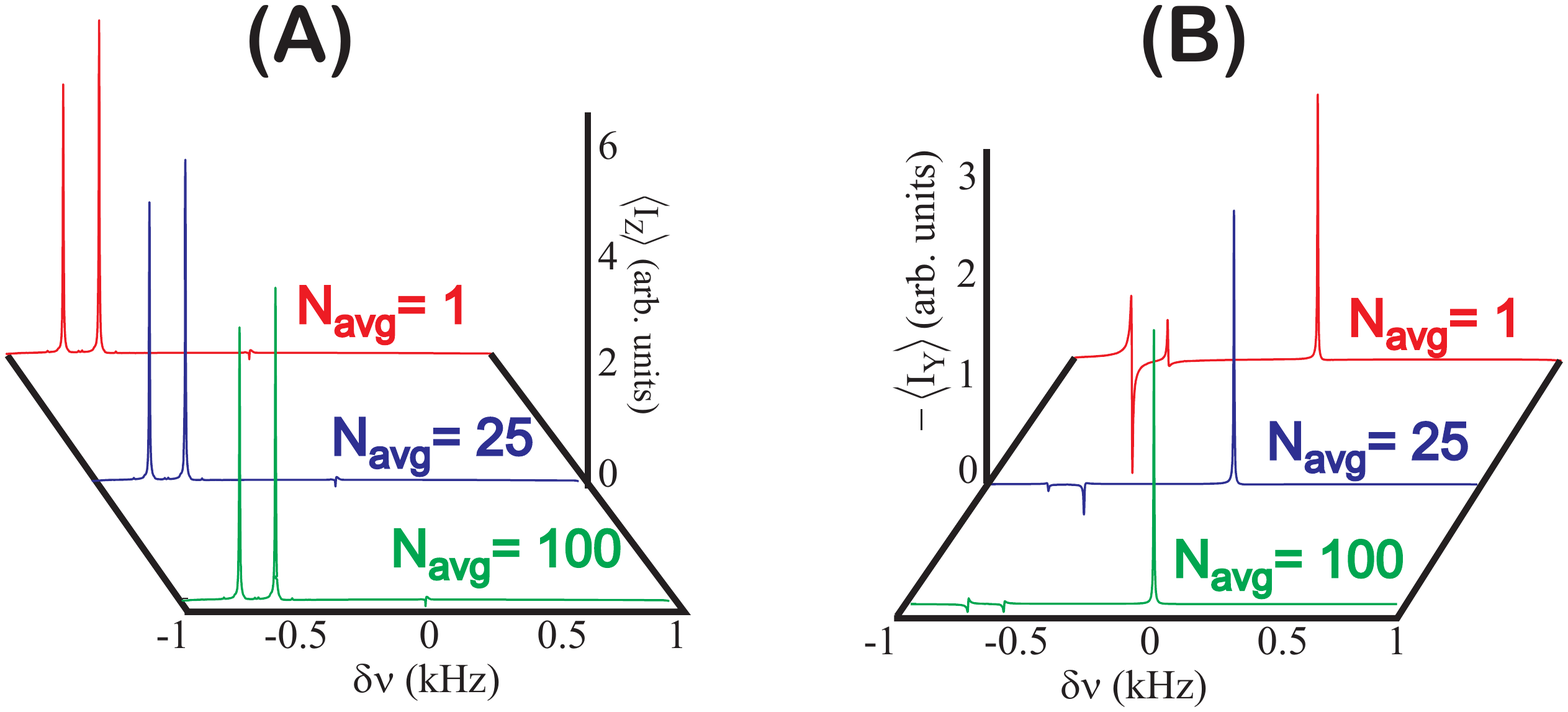}
{\small{\caption{Experimental stacked plots of the (A) $\widehat{z}$-magnetization weighted spectra and the (B) excitation spectra after averaging over $N_{avg}=1$ (red), $N_{avg}=25$ (blue), and $N_{avg}=100$ (green) p-DANTE sequences [same sequences used in Figs. \ref{fig:figure4}(C) and \ref{fig:figure4}(D)] applied to the acetone, DMSO, and water solution used in Fig. \ref{fig:figure7}.  From Figs. \ref{fig:figure4}(C) and \ref{fig:figure4}(D), only the water resonance at $\delta\nu=0$ Hz should be efficiently excited.  From Fig. \ref{fig:figure6}(B) the water resonance $(\delta\nu=0)$ is efficiently excited, and the amount of excitation at the acetone and DMSO resonances decreases with averaging over more p-DANTE sequences.  In \ref{fig:figure8}(A), the $\widehat{z}-$ magnetization weighted spectra are shown, illustrating that both the acetone and DMSO magnetization lie mostly along the $\widehat{z}$-direction after application of the p-DANTE sequences.\label{fig:figure8}}}}
\end{figure}

  The averaged excitation [Fig. \ref{fig:figure8}(B)] and $\widehat{z}-$magnetization weighted spectra [Fig. \ref{fig:figure8}(A)] for the DMSO/acetone/water solution was obtained using up to one hundred different p-DANTE sequences, and the results are shown for $N_{avg}=1$ (red curve), $N_{avg}=25$ (blue curve), and $N_{avg}=100$ (green curve) in Figure \ref{fig:figure8}.  The p-DANTE sequences used in Fig. \ref{fig:figure8} were the same used in the theoretical calculations shown in Figs. \ref{fig:figure4}(C) and \ref{fig:figure4}(D), where the $k^{th}$ delay used in $p^{th}$ experiment was given by:
\begin{eqnarray}
\tau^{p}_{k}=\frac{46.77\text{ms}}{29-\frac{1}{2\sqrt{2}}\left(1-\csc\left(\frac{\pi}{f_{p}}\right)\sin\left(\frac{59\pi}{f_{p}}\right)\right)}\left(1+\frac{1}{\sqrt{2}}\cos\left(\frac{2\pi k}{f_{p}}\right)\right)
\end{eqnarray}
which ensured that the average delay, $\frac{1}{29}\sum_{k=1}^{29}\tau_{k}^{p}\approx \frac{1}{|\delta\nu_{\text{DMSO,water}}|}=1.6$ ms, was the same as the delay used in the DANTE sequence shown in Figs. \ref{fig:figure7}(B) and \ref{fig:figure7}(C).  As in Figs. \ref{fig:figure4}(C) and \ref{fig:figure4}(D), the water ($\delta\nu=0$ Hz) was maximally excited whereas the averaged excitation at the acetone and DMSO resonances decreased upon averaging over different p-DANTE sequences.  Similarly, the $\widehat{z}$-magnetization weighted spectra indicated that the acetone and DMSO magnetization remained mostly about the $\widehat{z}$-axis after application of the p-DANTE sequence, whereas there was little $\widehat{z}$-magnetization at the water resonance.

\section{Conclusions}
In this work, average Hamiltonian theory (AHT) was used to calculate the effective propagators for the both the DANTE [Fig. \ref{fig:figure1}(A)] and pseudorandom-DANTE or p-DANTE [Fig. \ref{fig:figure1}(B)] sequences.  It was found that an AHT description the DANTE sequence is valid when $\theta=\frac{\pi}{60}$ and for total pulse flip-angles of  $\Theta=N\theta\leq \frac{5\pi}{9}$ over all frequencies [Fig. \ref{fig:figure3}].  The validity of the AHT description was also found to depend upon the spin's resonance frequency, $\nu$, and an AHT description of DANTE works well for frequencies in the range $\delta\nu\gg\frac{2\theta}{5\tau_{t}\Theta}$ and $\delta\nu\ll \frac{2\theta}{5\tau_{t}\Theta}$ where $\delta\nu=\text{min}\left[\left|\nu-\frac{n}{\tau_{t}}\right|\right]$ is the smallest frequency difference between $\nu$ and the nearest resonance of the DANTE sequence, $\frac{n}{\tau_{t}}$ where $n$ is an integer.  Understanding the limitations of an AHT description for DANTE enabled us to develop an AHT description of the p-DANTE sequences [Fig. \ref{fig:figure1}(B)] where the delays and phases of the pulses are modulated in concert throughout the sequence.  This modulation of delays and phases breaks the periodicity of the DANTE sequence and enables the p-DANTE sequence to excite spins at a single frequency, $\nu_{0}$.  The ability to use an AHT description for the DANTE and p-DANTE sequences might also provide additional insights into other selective pulse sequences, since any shaped pulses can be cast into a DANTE-like description\cite{Shinnar87,Shinnar89,Shinnar89a}.  While the excitation and $\widehat{z}-$magnetization profiles for a single p-DANTE sequence are not particularly clean, i.e., small excitations exist at many frequencies, averaging over different p-DANTE sequences helps to ''clean-up" the excitation profiles so that only a baseline excitation exists everywhere except at $\nu_{0}$, which is excited.  Experimental demonstrations [Fig. \ref{fig:figure6} and Fig. \ref{fig:figure8}] of the p-DANTE sequences were found to be in good agreement with theoretical predictions.

 For future work, determining the existence of an optimal set of p-DANTE sequences that generate the "cleanest" excitation profiles using the smallest number of p-DANTE sequences will be investigated.  Since the frequency selection in p-DANTE sequences is determined by correlating the pulse phases with the delays, the p-DANTE sequences could also be incorporated into ultrafast NMR\cite{Frydman02,Frydman03} techniques to selectively excite certain resonances in different parts of the sample volume. Furthermore, extending the AHT results obtained in this paper to coupled spin systems is currently underway, whereby a DANTE-like or p-DANTE-like sequences can be used to selectively excite a particular multiple-quantum spin transition.  The conditions under which such an AHT description can be applied in these systems are approximately the same as those found in this paper, since any subspace of two transitions can be described\cite{Feynman57} as an effective spin-1/2.  Coupling these techniques with ultra-fast NMR should enable the quick determination of all spin transitions in a given molecular system.

{\bf{Acknowledgments}} We would like to thank Alex Burum for a careful reading of this manuscript.  This work
was supported by a Camille and Henry Dreyfus New Faculty award, a Provost Research award and startup funds from the University of Miami.
\\
%\bibliographystyle{prstya}
%\bibliography{wallsbib}
%merlin.mbs 2010-03-15 4.21a (PWD, AO, DPC)
\end{document}